\documentclass[preprint,prl,amsmath,amssymb,showpacs]{revtex4}

\usepackage{graphicx}% Include figure files
\usepackage{dcolumn}% Align table columns on decimal point
\usepackage{bm}% bold math
\usepackage{overpic}

\begin{document}

%\preprint{}

\title{Superscatterer: Enhancement of scattering with complementary media}
\author{Tao Yang}
\author{Huanyang Chen}
\author{Xudong Luo}\thanks{To whom correspondence should be
addressed. \\ Email address: luoxd@sjtu.edu.cn}
\author{Hongru Ma}
\affiliation{Institute of Theoretical Physics, Shanghai Jiao Tong
University, Shanghai 200240, People's Republic of China}

\date{\today}

\begin{abstract}
Based on the concept of complementary media, we propose a novel
design which can enhance the electromagnetic wave scattering cross
section of an object so that it looks like a scatterer bigger than
the scale of the device. Such a ``superscatterer'' is realized by
coating a negative refractive material shell on a perfect electrical
conductor cylinder. The scattering field is analytically obtained by
Mie scattering theory, and confirmed by full-wave simulations
numerically. Such a device can be regarded as a cylindrical concave
mirror for all angles.

\end{abstract}


\maketitle

Recently great progress has been
made\cite{Leonhardt,Pendry2006,Greenleaf,Schurig,Alu,Cummer,concentrator,
Jacob,Cai,Chenhy,HChen,Ruan,Zhang} in manipulating the
electromagnetic (EM) fields by means of metamaterials. By employing
the coordinate transformation approach proposed by Leonhardt
\cite{Leonhardt} and Pendry {\it et al.} \cite{Pendry2006}, various
exciting functional EM devices have been reported
\cite{Schurig,concentrator,Chenhy}. This methodology provides a
clear geometric picture of those designed devices,  and the
permittivity and permeability tensors of designed functional
materials can be derived from coordinate transformations directly.
On the other hand, Mie scattering theory \cite{HChen,Ruan,Yaghjian}
provided an analytic approach to quantitatively analyze the
scattering properties of EM fields, and the aforementioned
functional devices can also be realized by choosing different scalar
transformation functions. The combination of the geometric and the
analytic approaches may give novel and surprising results. For
example, in contrast to invisibility cloak, we may design an EM
transformation media device to enlarge the scattering cross-section
of a small object. In this way the object is effectively magnified
to a size larger than the object plus the device so that it is much
easer for EM wave detection, which we refer hereafter as a
superscatterer.

In this Letter, we propose a generalized technique to achieve such a
superscatterer. In the quasistatic limit, Nicorovici {\it et al.}
\cite{Nicorovici} had demonstrated that the properties of a coated
core can be extended beyond the shell into the matrix. It is called
a {\it partially-resonant} system in which at least one dielectric
constant of three-phase composite structure is negative. Moreover,
by using the negative refracting material (NRM)
\cite{Veselago,Pendry2000}, Pendry and Ramakrishna \cite{Pendry2002}
showed how to image an object by perfect cylindrical lens based on
the concept of complementary media. Here we limit our discussion on
the 2D case and prove that a special magnified image acts as a real
object for EM wave detection.

\begin{figure}
\includegraphics[width=3.50in]{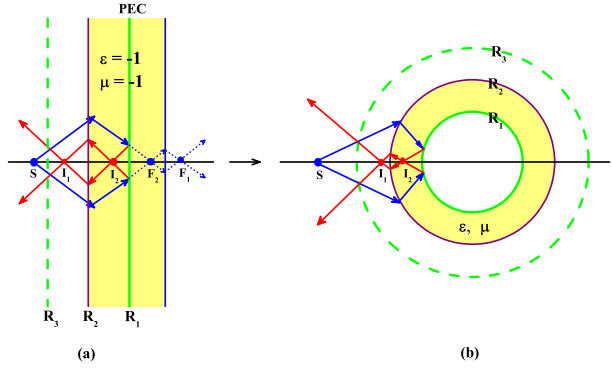}
\caption{\label{light} (a) The schematic demonstration  of the
behavior of a beam propagates in a complementary media and PEC
boundary. (b) The behavior is extended to the 2D case, where a
superscatterer is formed with its effective size shown by the dashed
line.}
\end{figure}

Figure \ref{light} shows a heuristic model stemmed from the
concept of complementary media \cite{Pendry2002}. In Fig.
\ref{light}(a), a slab of metamaterial with $\epsilon=\mu=-1$
images point source $S$ at $F_2$ inside and $F_1$ outside the
metamaterial if the distance between source and slab is less than
the slab thickness. Now a perfect electrical conductor(PEC)
boundary is set at $R_1$, so the light propagating in the slab is
completely reflected by it and focuses at $I_2$ inside and $I_1$
outside the slab. Since the slab and the vacuum with the same
thickness are complementary to each other, it looks as if the
medium between $R_3$ and $R_1$ is moved, and the PEC at $R_1$ is
virtually shifted to $R_3$ and takes effect as a real PEC
boundary. The picture can be extended to the 2D case as shown in
Fig. \ref{light}(b), in which $z$ axis is perpendicular to the
paper. In the 2D case, the $R_1$, $R_2$ and $R_3$ represent the
radius of the inner region, the outer radius of the cylindrical
annulus (metamaterial shell) and the effective radius of the
virtual cylinder, respectively. In order to move effectively the
inner PEC boundary at $r=R_1$ to the surface of the virtual
cylinder, where the radius is $r=R_3>R_2$, both permittivity and
permeability tensors in the metamaterial shell should be selected
properly. The Mie scattering theory is a powerful tool to
accomplish the goal.

We consider a transverse-electric (TE) polarized EM incident field
with harmonic time dependence $\exp{(-iwt)}$. The coordinate system
is the cylindrical coordinate coinciding with the cylinder we
considered. In this coordinate system the permittivity and
permeability tensors can be put in the following general form
\begin{equation}
\overline{\overline\epsilon}=\epsilon_r(r)\hat{r}\hat{r}
+\epsilon_\theta(r)\hat{\theta}\hat{\theta}
+\epsilon_z(r)\hat{z}\hat{z}, \quad\quad
\overline{\overline\mu}=\mu_r(r)\hat{r}\hat{r}
+\mu_\theta(r)\hat{\theta}\hat{\theta} +\mu_z(r)\hat{z}\hat{z}.
\end{equation}
The EM fields in the homogeneous material region ($r>R_2$) are well
known and can be expressed by the superposition of cylindrical
functions. The wave equation of $E_z$ in the shell is written as
\begin{equation}\label{eq2}
\frac{1}{\epsilon_z} \frac{1}{r} \frac{\partial}{\partial r}
\left(\frac{r}{\mu_\theta} \frac{\partial E_z}{\partial r}\right) +
\frac{1}{\epsilon_z} \frac{1}{r^2} \frac{\partial}{\partial\theta}
\left(\frac{1}{\mu_r} \frac{\partial E_z}{\partial_\theta }\right) + k_0^2E_z
=0,
\end{equation}
where $k_0$ is the wave vector of the EM wave in vacuum. Now we
introduce a new coordinate system $(f(r),\theta,z)$, in which $f(r)$
is a continuous and piecewise differentiable function of the
original radial coordinate $r$, the wave equation of field
$\tilde{E}_z(f(r),\theta)=E_z(r,\theta)$ is transformed to
\begin{eqnarray}\label{eq3}
\frac{1}{\epsilon_z}\frac{1}{r}f'\frac{\partial}{\partial f}
\left(\frac{r}{\mu_\theta}f'\frac{\partial \tilde{E}_z}{\partial
f}\right)
+\frac{1}{\epsilon_z}\frac{1}{r^2}\frac{\partial}{\partial\theta}
\left(\frac{1}{\mu_r}\frac{\partial \tilde{E}_z}{\partial_\theta
}\right)+k_0^2\tilde{E}_z=0,
\end{eqnarray}
where $f'(r)$ denotes ${\text d} f(r)/{\text d}r$. By taking the
components of permittivity and permeability tensors as
\cite{Yaghjian},
\begin{eqnarray}\label{eq4}
\begin{aligned}
\frac{\epsilon_r}{\epsilon_0}=\frac{\mu_r}{\mu_0}
=\frac{f(r)}{r}\frac{1}{f'(r)},
\\
\frac{\epsilon_\theta}{\epsilon_0}=\frac{\mu_\theta}{\mu_0}
=\frac{r}{f(r)}f'(r),
\\
\frac{\epsilon_z}{\epsilon_0}=\frac{\mu_z}{\mu_0}
=\frac{f(r)}{r}f'(r),
\end{aligned}
\end{eqnarray}
where $\epsilon_0$ and $\mu_0$ are the vacuum  permittivity and
permeability, the Eq. \eqref{eq3} can be solved by separation of
variables  $\tilde{E}_z=R(f)\Theta(\theta)$. The solution of $R(f)$
and $\Theta(\theta)$ are just the $m$th-order Bessel functions and
$\exp(im\theta)$, in which $m$ are integers.

With the above analysis, the electric fields in each domain are
expressed as
\begin{eqnarray}\label{eq9}
E_z(r,\theta)=\left\{
\begin{array}{lr}
0, & \quad  r<R_1, \\
\sum_m{\left(\alpha_m^iJ_m(k_0f(r))
+\alpha_m^sH_m^{(1)}(k_0f(r))\right)\exp(im\theta)},
&\quad  R_1<r<R_2, \\
\sum_m{\left(\beta_m^iJ_m(k_0r)
+\beta_m^sH_m^{(1)}(k_0r)\right)\exp(im\theta)}, & \quad r>R_2.
\end{array}\right.
\end{eqnarray}
where $J_m$ and $H_m^{(1)}$ are the $m$th-order Bessel function and
Hankel function of the first kind, respectively. By means of the
orthogonality of $\exp(im\theta)$ and the continuity of $E_z$ and
$H_\theta$ at two interfaces ($r=R_1$ and $r=R_2$), we obtain the
linear relationships between the coefficients as follows,
\begin{subequations}\label{eq10}
\begin{eqnarray}
\alpha_m^iJ_m(k_0f(R_1))+\alpha_m^sH_m^{(1)}(k_0f(R_1))&=&0, \\
\alpha_m^iJ_m(k_0f(R_2))+\alpha_m^sH_m^{(1)}(k_0f(R_2))
&=&\beta_m^iJ_m(k_0R_2)+\beta_m^sH_m^{(1)}(k_0R_2
),\\
\frac{R_2}{f(R_2)}\left[\alpha_m^iJ'_m(k_0f(R_2))
+\alpha_m^sH_m^{(1)}{'}(k_0f(R_2))\right]&=&
\beta_m^iJ'_m(k_0R_2)+\beta_m^sH_m^{(1)}{'}(k_0R_2 ),
\end{eqnarray}
\end{subequations}
where the prime denotes differentiation with respect to the entire
argument of Bessel functions. By imposing the boundary condition
of $f(r)$, $f(R_2)=R_2$, we get the solutions of Eqs.\eqref{eq10}
as follows,
\begin{eqnarray}\label{eq11}
\frac{\alpha_m^s}{\alpha_m^i} =\frac{\beta_m^s}{\beta_m^i}
=-\frac{J_m(k_0f(R_1))}{H_m^{(1)}(k_0f(R_1))}, \quad \quad m=0,\pm
1, \pm2,\dots .
\end{eqnarray}
This is the exact solution for scattering matrix.

We can draw some interesting conclusions from Eqs. \eqref{eq11}. If
$f(r)$ is a monotonic function and $f(R_1)=R_3>R_2$, the material of
the shell must be NRM from Eqs.\eqref{eq4}, and the fields at point
$(r,\theta,z)$ in inner annulus $R_1<r<R_2$ are equal to those at
point $(f(r),\theta,z)$ in outer region $R_2<r<R_3$. Moreover, in
the region $r>R_3$, the scattering fields are exactly equal to those
scattered by a PEC cylinder with radius $R_3$, as if whole
cylindrical annulus of $R_1<r<R_3$ is moved and the PEC boundary at
$r=R_1$ is magnified and shifted to $r=R_3$. Here, the cylinder with
radius $f(R_1)$ is called a virtual cylinder. It is just the
expected result in Fig. \ref{light}(b).
\begin{figure}[h]
\includegraphics[width=2.50in]{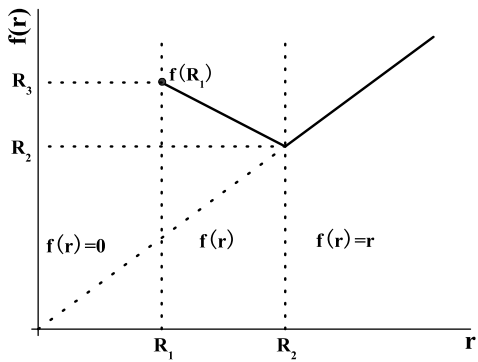}
\caption{\label{fr} A simple function $f(r)$ satisfies the condition
of Fig.\ref{light}(b)}
\end{figure}

For simplicity, we choose a linear function for $f(r)$,
\begin{equation}\label{eq1}
f(r)= \left\{
\begin{array}{lr}
b_0(R_2-r)/(R_2-R_1)+R_2,  & \quad R_1 <r < R_2, \\
r, & \quad\quad\quad\; r>R_2 .
\end{array}\right.
\end{equation}
Here $b_0$ is an important parameter to determine the magnification
factor. It should be noted that both cloak and concentrator can also
be obtained in the same manner. For example, it is a perfect cloak
when $b_0=-R_2$, and it becomes an imperfect cloak when
$R_1-R_2>b_0>-R_2$, or a concentrator when $b_0>R_1-R_2$ and
$f(r)=rf(R_1)/R_1$ in the region $r<R_1$. However, the case $b_0>0$,
which provides a folded geometry \cite{Leonhardt2006,Milton}, is
less discussed. In fact, with the help of the geometric picture of
the concentrator, it means a big cylinder with radius $b_0+R_2$ is
compressed into a small cylinder with radius $R_1$, and the gap
between $r=R_1$ and $r=b_0+R_2$ has to be filled by a pair of
complementary media: One is the vacuum in $R_2<r<b_0+R_2$ and the
other is the NRM shell in $R_1<r<R_2$. This device scatters the same
fields as the uncompressed cylinder, which extends beyond the shell,
so we call it ``superscatterer''.

\begin{figure}
\begin{overpic}[scale=0.6,bb=0 0 336 228]{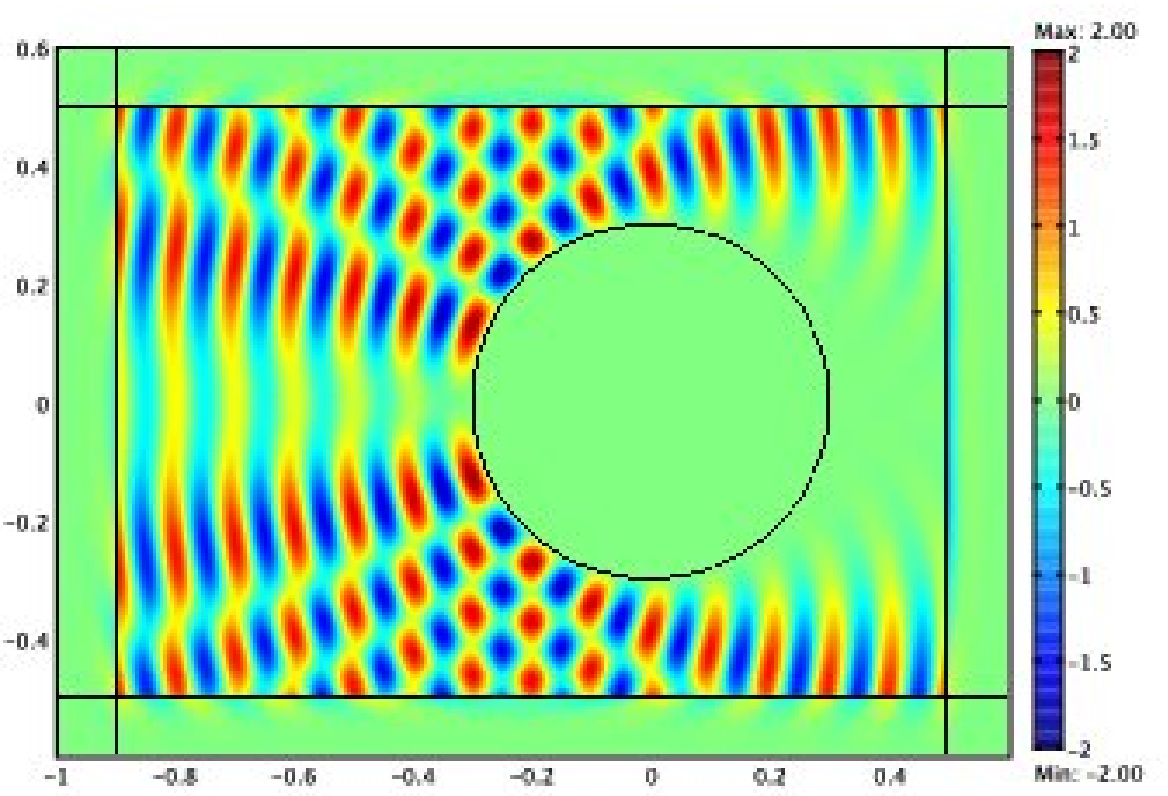}
\put(72,14){\bf(a)}
\end{overpic}
\begin{overpic}[scale=0.6,bb=0 0 336 228]{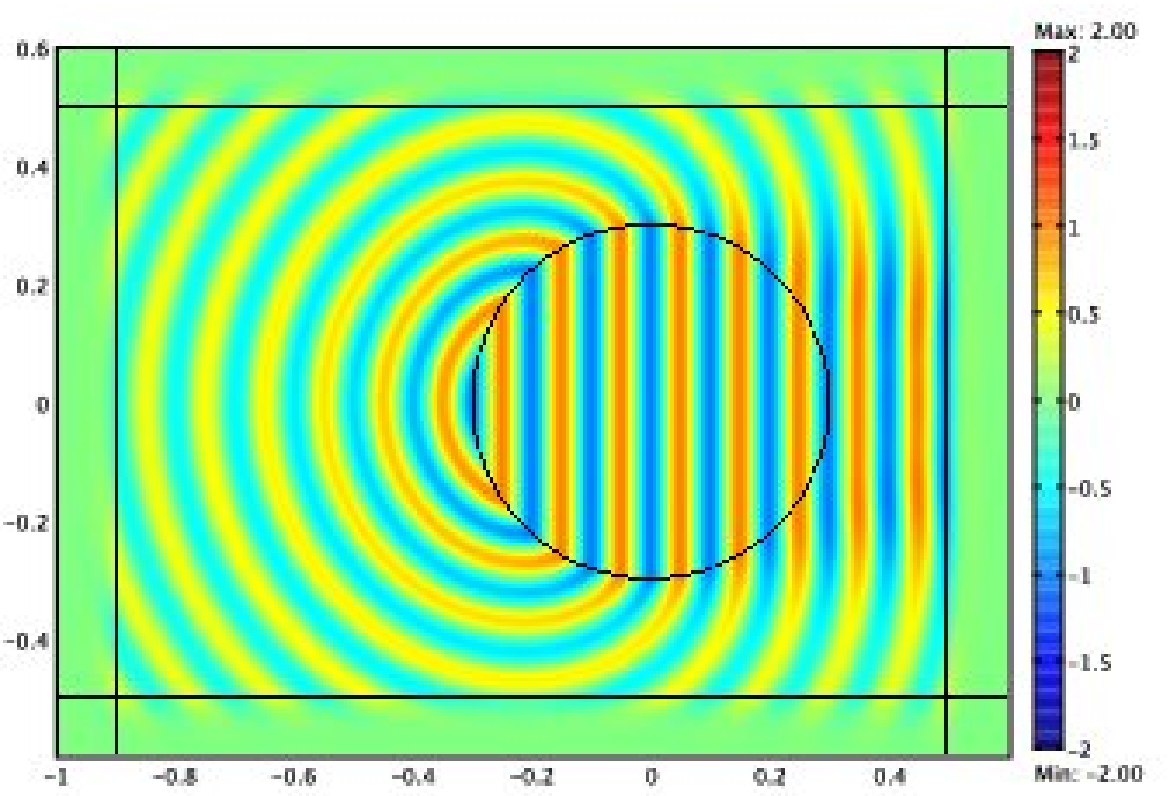}
\put(72,14){\bf(b)}
\end{overpic}
\begin{overpic}[scale=0.6,bb=0 0 336 228]{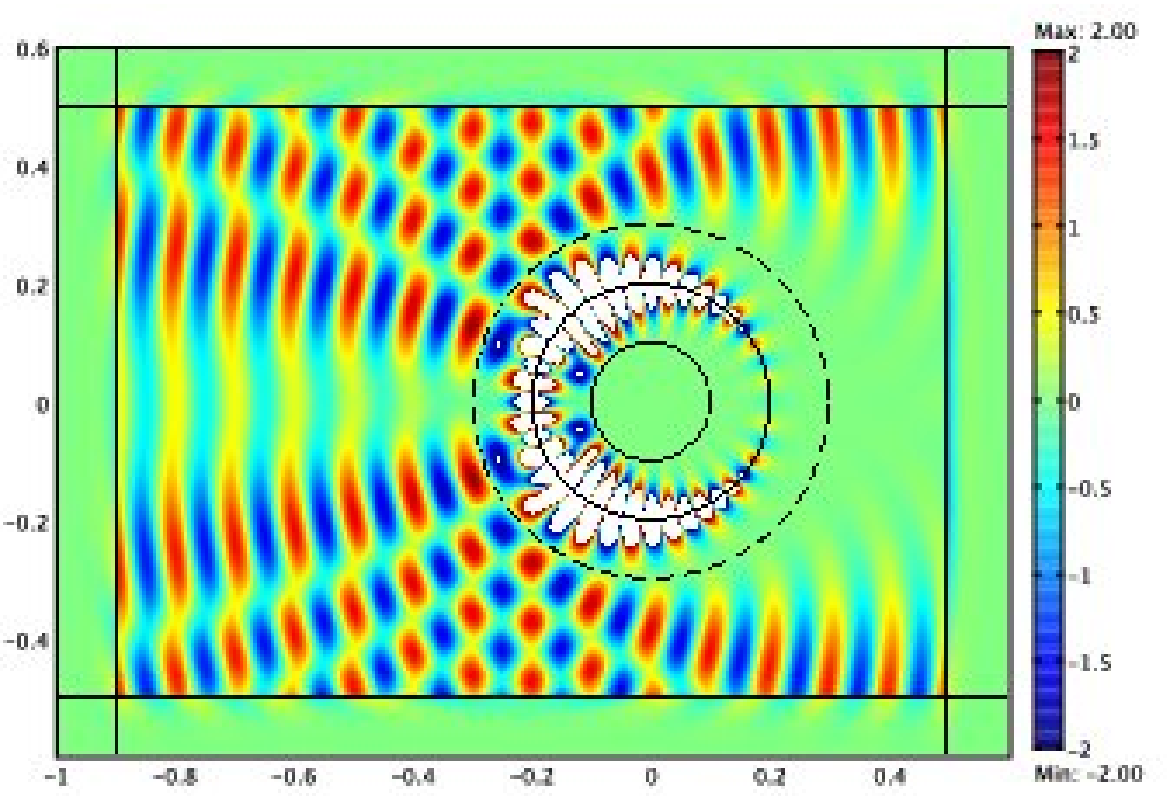}
\put(72,14){\bf(c)}
\end{overpic}
\begin{overpic}[scale=0.6,bb=0 0 336 228]{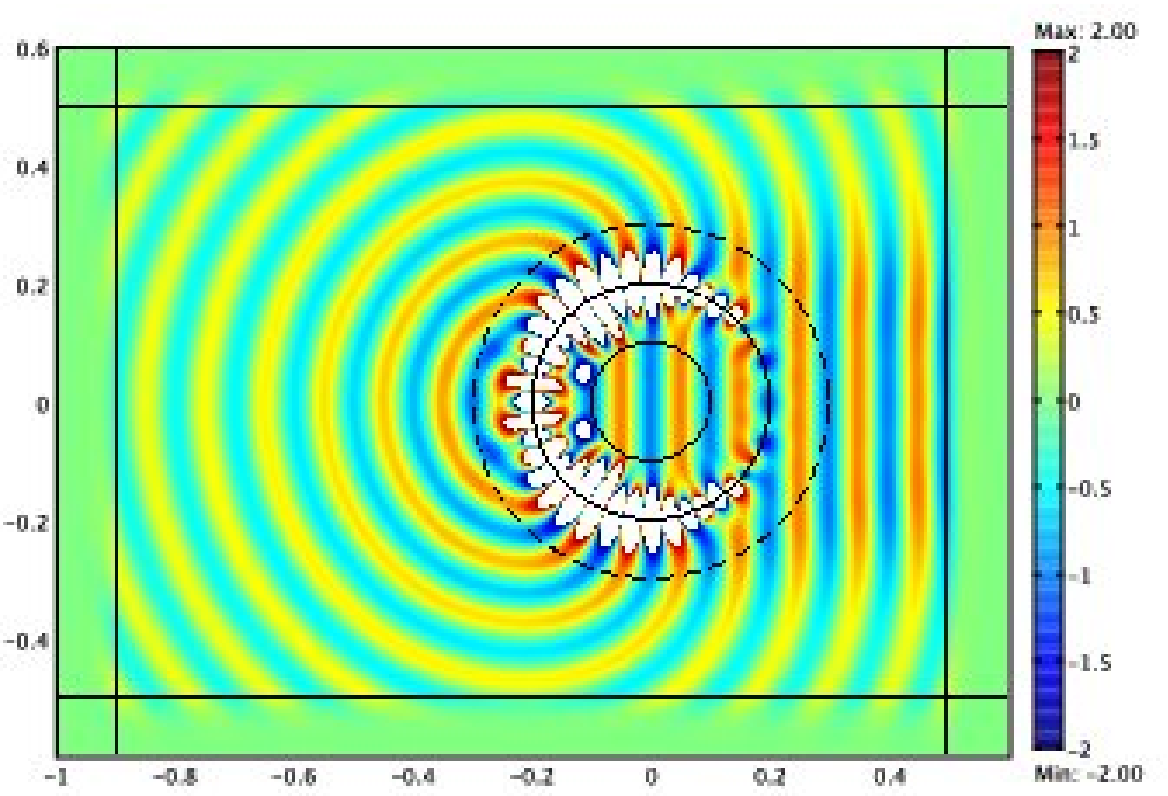}
\put(72,14){\bf(d)}
\end{overpic}
\caption{\label{planewave} Snapshot of the total and scattering
electric field. (a) - (b) The total and scattering electric fields
induced by PEC cylinder with radius $R_3=0.3$m, respectively. (c) -
(d) The total and scattering electric fields induced by the designed
device (the radius of virtual cylinder is  $0.3$m), respectively.}
\end{figure}

Next, we present the patterns of electric field calculated by
finite element solver of the Comsol Multiphysics software package.
In Fig. \ref{planewave}, the plane wave is normal incident from
left to right with frequency $3$ GHz and unit amplitude, and the
inner and outer radii of the shell are $R_1=0.1$m and $R_2=0.2$m,
respectively. When $b_0$ is equal to $0.1$m and the function
$f(r)$ is taken as in Eq. \eqref{eq1}, the radius of the virtual
cylinder becomes $f(R_1)=b_0+R_2=0.3$m. The ranges of the
components of $\overline{\overline\epsilon}$ and
$\overline{\overline\mu}$ are taken as follows: $\epsilon_r, \mu_r
\in [-3,-1]$, $\epsilon_\theta, \mu_\theta \in [-1,-\frac{1}{3}]$,
and $\epsilon_z, \mu_z \in [-3,-1]$. A tiny absorptive imaginary
part ($\sim 10^{-5}$) is added to $\overline{\overline\epsilon}$
and $\overline{\overline\mu}$ due to the inevitable losses of the
NRM.

Fig. \ref{planewave}(a) and \ref{planewave}(b) are snapshots of the
total electric field and scattering electric field induced by a PEC
cylinder with radius $R_3=0.3$m, respectively; Fig.
\ref{planewave}(c) and \ref{planewave}(d) are those fields induced
by the superscatterer. Comparing the patterns of electric field in
the region $r>R_3$, one can find they are almost equivalent. Here,
the bounds of the amplitude of electric field in Fig.
\ref{planewave} set from $-2$ to $2$ for clarity. The white flecks
in the region of the complementary media show the regions where the
values of fields exceed the bounds. The highest value of the field
in the flecks is about $10^2$, which comes from the dominative
high-$m$ modes with the factor $J_m(k_0R_3) H_m^{(1)}(k_0
f(r))/H_m^{(1)}(k_0R_3)$ in the scattering field.

We define the magnification factor $\eta$ as $f(R_1)/R_2$, which
is the ratio of the radius of the virtual cylinder to the real
size of this device. The components of permittivity and
permeability tensors are negative for the case of $\eta>1$. There
is no singularity for any finite $\eta$ as long as both $f(r)$ and
$f^{\prime}(r)$ are nonzero in $R_1\leq r \leq R_2$. For example,
when $\eta=20$, the ranges of the parameters are: $\epsilon_r,
\mu_r \in [-1.1,-0.052]$, $\epsilon_\theta, \mu_\theta \in
[-19,-0.90]$, and $\epsilon_z, \mu_z \in [-399,-19]$.

\begin{figure}
\begin{overpic}[scale=0.6,bb=0 0 336 228]{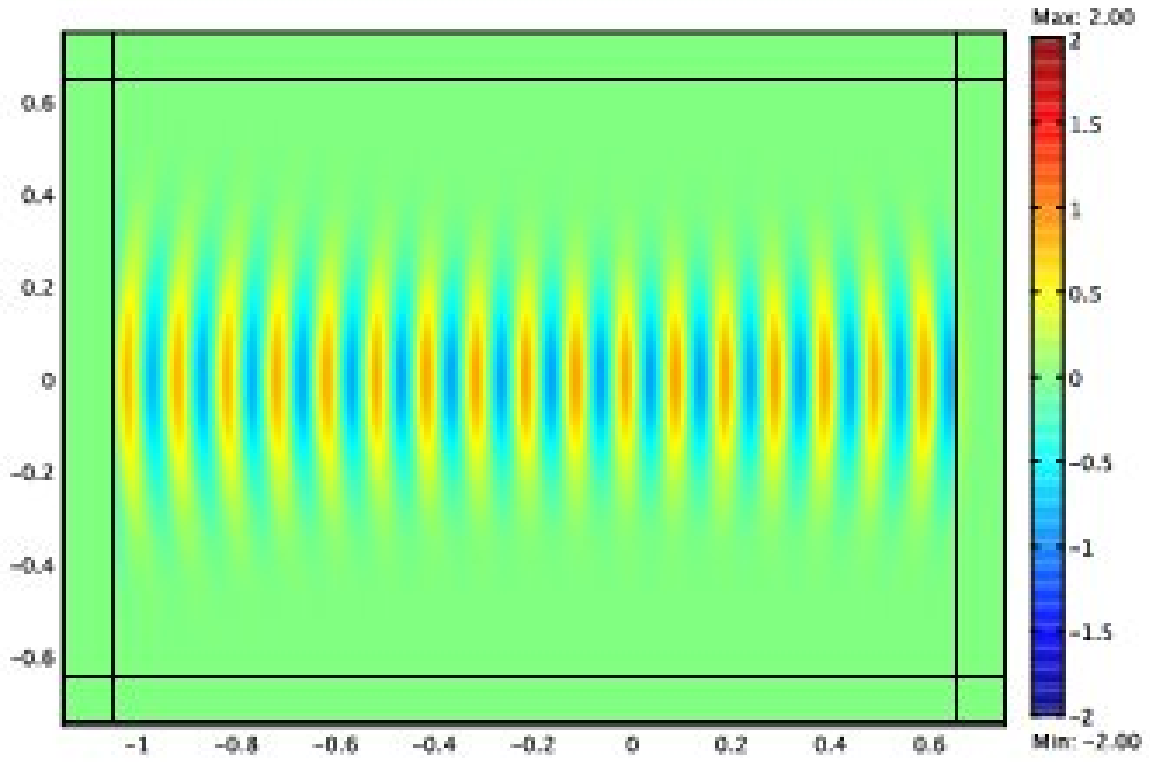}
\put(72,12){\bf(a)}
\end{overpic}
\begin{overpic}[scale=0.6,bb=0 0 336 228]{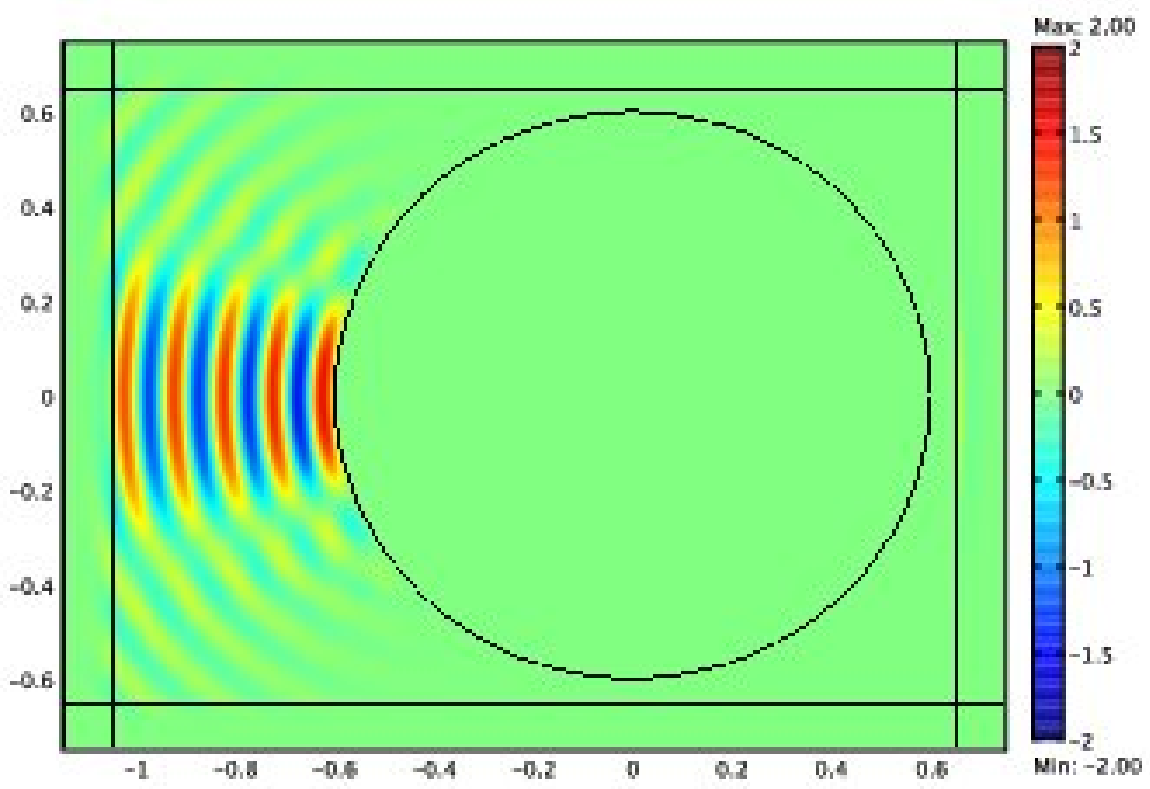}
\put(72,12){\bf(b)}
\end{overpic}
\begin{overpic}[scale=0.6,bb=0 0 336 228]{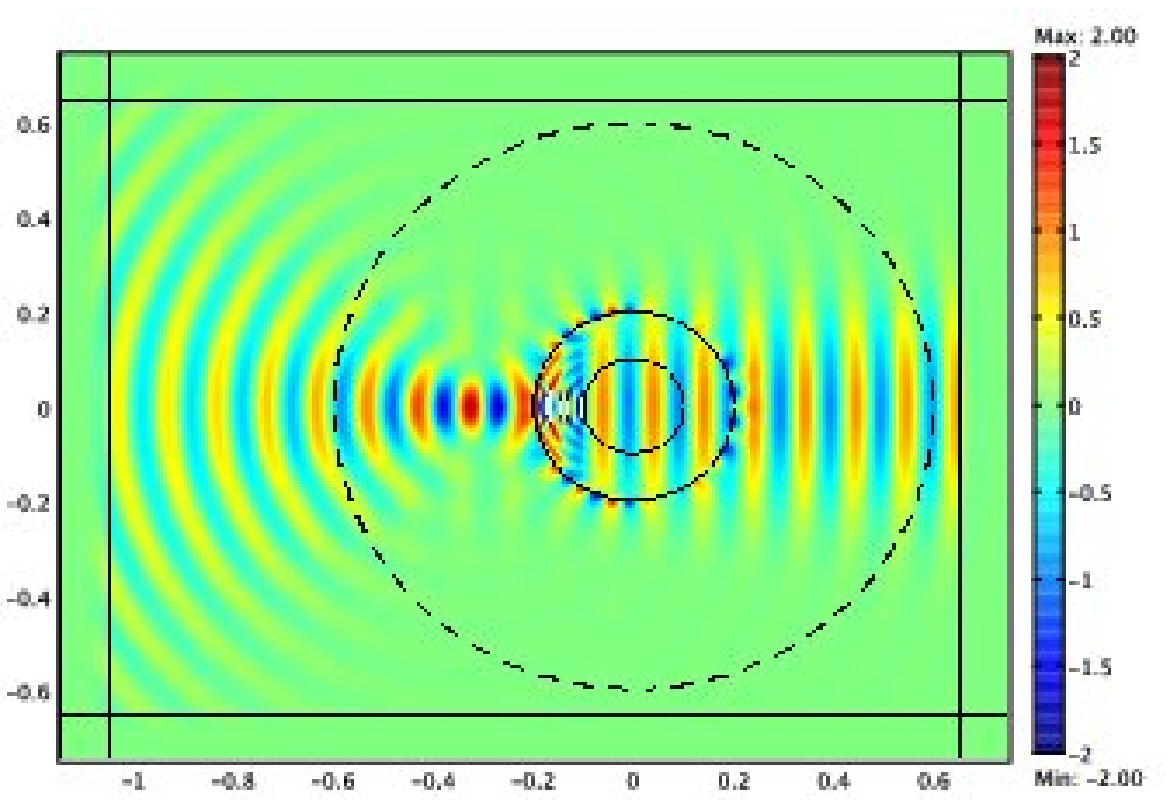}
\put(72,12){\bf(c)}
\end{overpic}
\begin{overpic}[scale=0.6,bb=0 0 336 228]{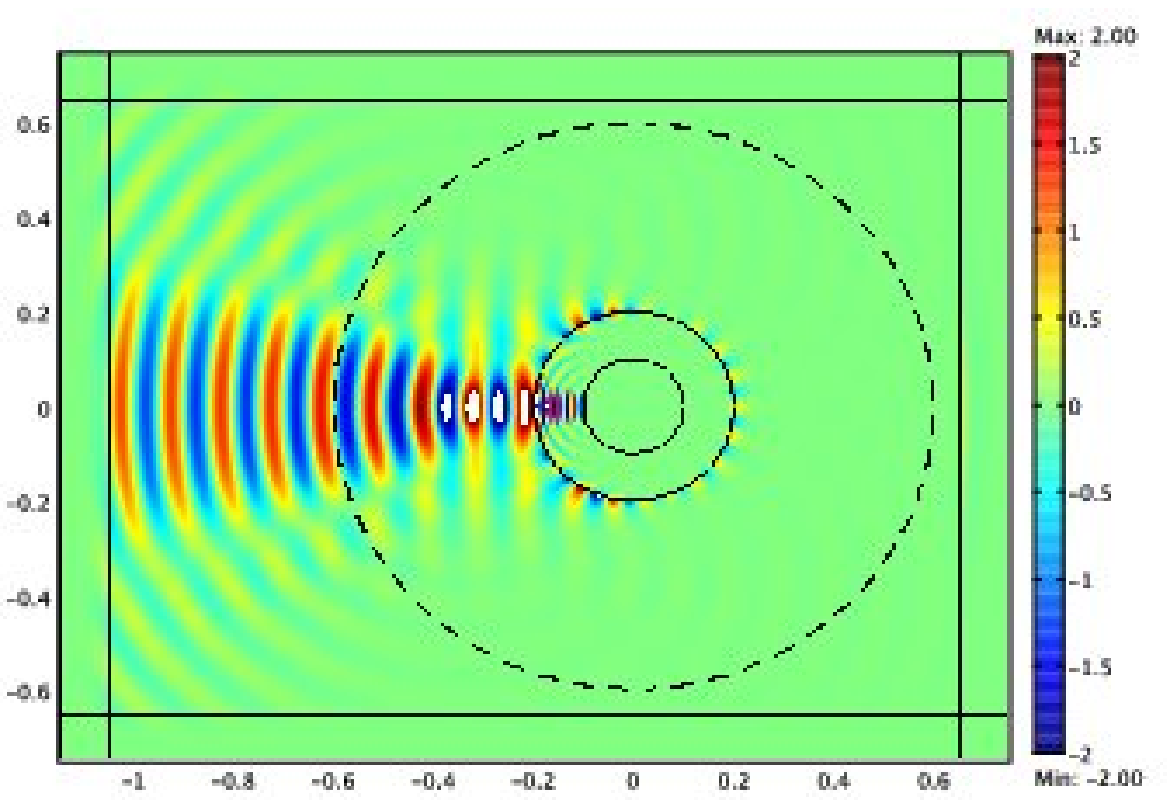}
\put(72,12){\bf(d)}
\end{overpic}
\caption{\label{compare} (a) A Gaussian beam in free space. (b)
The total electric field induced by PEC cylinder with radius
$0.6$m. (c) - (d) The scattering and total fields by the
superscatterer (the radius of the virtual cylinder is $0.6$m),
respectively.}
\end{figure}

Moreover, the device in Fig. \ref{light}(b) can be regarded as a
cylindrical concave mirror for all angles, which can not be realized
by ordinary media. From the theory of geometrical optics, when a
plane wave incidents on it, the paraxial beams focus at
$r=f(R_1)/2=(b_0+R_2)/2$. Here, a full-wave simulations with COMSOL
Multiphysics are used to test the focus behavior of a Gaussian beam
propagated from left to right with unit amplitude. Fig.
\ref{compare}(a)-(d) give the snapshots of the Gaussian beam in free
space, the total electric field induced by PEC cylinder with radius
$0.6$m, and the scattering and total electric fields induced by the
superscatterer (here $R_1=0.1$m, $R_2=0.2$m and $b_0=0.4$m),
respectively. The scattering electric field in Fig. \ref{compare}(c)
shows the focus is approximately at $r=0.3$m since Gaussian beam is
close to a paraxial beam. Here, it is worthy noting that the
superscatterer becomes a cylindrical concave mirror for all angles
if and only if there is $b_0>R_2$. In principle, if we make the
radius $f(R_1)$ of virtual cylinder to infinity by adjusting the
function $f(r)$, the cylindrical concave mirror becomes a plane
mirror for all angles, so that any parallel light incidents to the
superscatterer will be reflected straight back along the incident
path.

In conclusion, we demonstrated the properties of a
``superscatterer'' in terms of PEC boundary and properly
complementary media. This kind of functional devices might be
important in EM detection. Similar concept can be extended to the
case of non PEC boundary and three dimension.

The work is supported by the National Natural Science Foundation of
China under grant  No.10334020  and in part by the National Minister
of Education Program for Changjiang Scholars and Innovative Research
Team in University.

\end{document}